\documentclass[]{spie}

\usepackage[]{graphicx}
\usepackage{amsmath, amssymb, amsfonts}
\usepackage[english]{babel}
\usepackage[]{epstopdf}
\usepackage[font={small,it}, margin = 0.5cm]{caption}

\title{	Deformable mirror interferometric analysis for the direct imagery of exoplanets} 

\author{Johan Mazoyer\supit{a}, Rapha\"e{}l Galicher\supit{a}, Pierre Baudoz\supit{a}, Patrick Lanzoni\supit{b}, Fr\'e{}d\'e{}ric Zamkotsian\supit{b} and G\'e{}rard Rousset\supit{a}
\skiplinehalf
\supit{a} LESIA, Observatoire de Paris, CNRS, UPMC Paris 6 and Denis Diderot Paris 7, Meudon, France; \\
\supit{b}Laboratoire d'Astrophysique de Marseille, CNRS, Aix-Marseille Univ., Marseille, France
}

\authorinfo{Further author information: johan.mazoyer@obspm.fr}
 
\begin{document} 
  \maketitle

%%%%%%%%%%%%%%%%%%%%%%%%%%%%%%%%%%%%%%%%%%%%%%%%%%%%%%%%%%%%% 
\begin{abstract}
Direct imaging of exoplanet systems requires the use of coronagraphs to reach high contrast levels ($10^{-8}$ to $10^{-11}$) at small angular separations (0.1$''$). However, the performance of these devices is drastically limited by aberrations (in phase or in amplitude, introduced either by atmosphere or by the optics). Coronagraphs must therefore be combined with extreme adaptive optic systems, composed of a focal plane wavefront sensor and of a high order deformable mirror. These adaptive optic systems must reach a residual error in the corrected wavefront of less than 0.1 nm (RMS) with a rate of 1 kHz. In addition, the surface defects of the deformable mirror, inherent from the fabrication process, must be limited in order to avoid the introduction of amplitude aberrations. 

An experimental high contrast bench has been developed at the Paris Observatory (LESIA). This bench includes a Boston Micromachine deformable mirror composed of 1024 actuators. For a precise analysis of its surface and performance, we characterized this mirror on the interferometric bench developed since 2004 at the Marseille Observatory (LAM).

In this paper, we present this interferometric bench as well as the results of the analysis. This will include a precise surface characterization and a description of the behavior of the actuators, on a 10 by 10 actuator range (behavior of a single actuator, study of the cross-talk between neighbor actuators, influence of a stuck actuator) and on full mirror scale (general surface shape).
\end{abstract}

%>>>>list of keywords


\section{Introduction}

Direct imaging of exoplanets requires the use of high-contrast imaging techniques among which coronagraphy. These instruments diffract and block the light of the star and allow us to observe the signal of a potential companion. However, these instrument are drastically limited by aberrations, introduced either by the atmosphere or by the optics themselves. The use of deformable mirrors (DM) is mandatory to reach the required performance. The THD bench (french acronym for very high-contrast bench), located in the Paris Observatory, in Meudon, France, uses coronagraphy techniques associated with a Boston Micromachines DM\cite{Bifano11}. This DM is a Micro-Electro-Mechanical Systems (MEMS), composed of 1024 actuators. In March 2013, we brought this DM in Laboratoire d\`{}Astrophysique de Marseille (LAM), France, where we studied precisely the performance and defects of this DM on the interferometric bench of this laboratory. The result of that study, conducted in collaboration with F. Zamkotsian et P. Lanzoni, from LAM are presented here.

We first describe the MEMS DM, the performance announced by Boston Micromachines and its assumed state before this analysis (Section~\ref{sec:DM_Boston}). In the same section, we also present the interferometric bench at LAM. The results of this analysis are then presented in several parts. We first describe the analyzed DM overall shape and surface quality (Sections~\ref{sec:forme_miroir} and \ref{sec:qualite_surface_DM}). We then analyze accurately the influence function of an actuator and its response to the application of different voltages (Section~\ref{sec:actionneur}), first precisely for one actuators and then extended to all the DM. Finally, special attention will be paid to the damaged actuators that we identified (Section~\ref{sec:dead_actioneur}). We will present several causes of dysfunction and possible solutions.

\section{The MEMS DM and the LAM interferometric bench}
\label{sec:DM_Boston}

\subsection{The MEMS DM: specifications and damaged actuators}

\begin{figure}[ht!]
 \parbox{0.65\textwidth}{ \centering
   \includegraphics[trim = 0.5cm 4.5cm 0.52cm 2.5cm, clip = true, width = 0.645\textwidth]{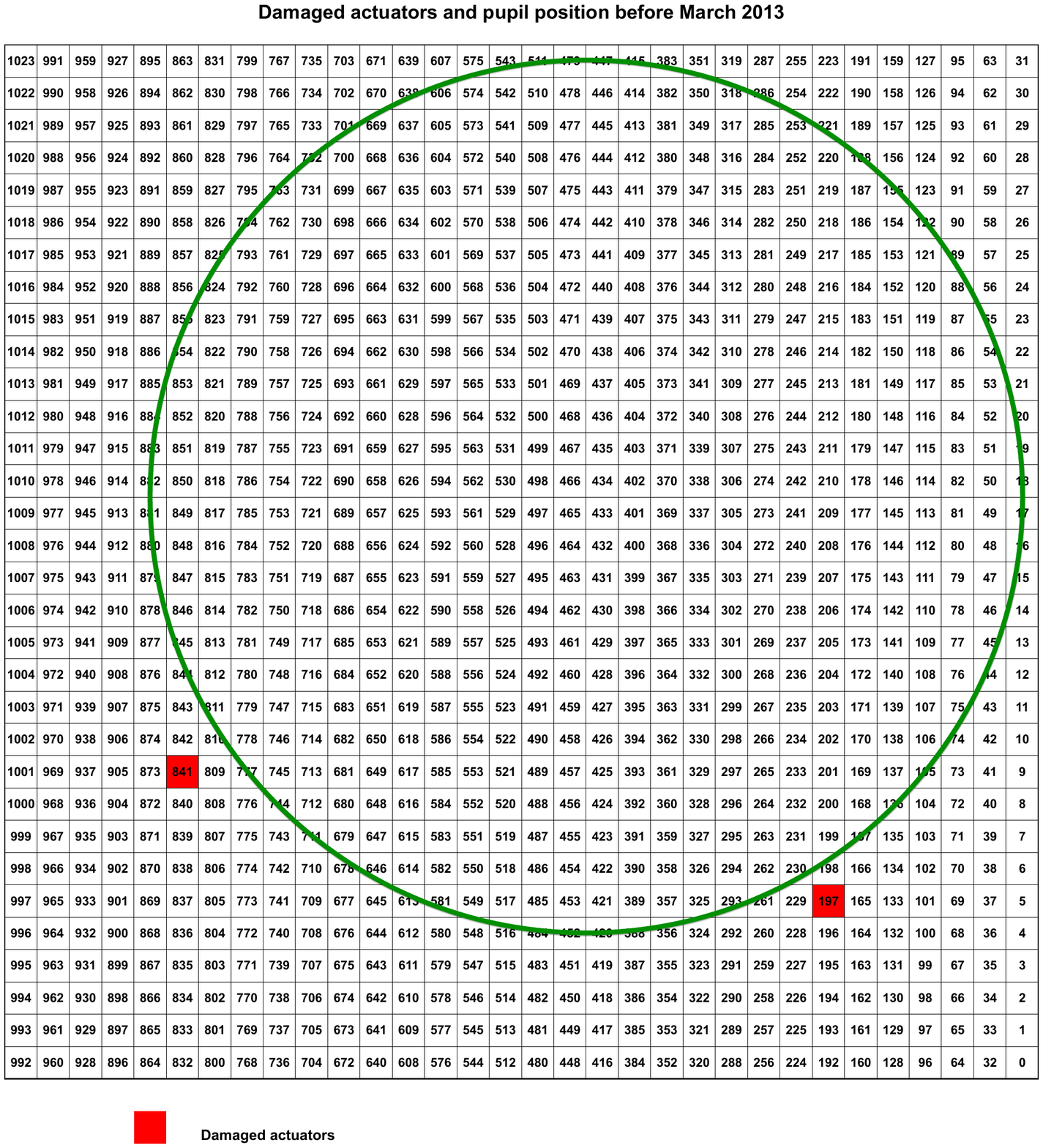}}
 \parbox{0.34\textwidth}{\centering \caption[beforepup] 
{ \label{fig:position_pupille_avant} Numbered actuators and position of the pupil on the DM before March 2013 (in green). The numbering starts at 0 in the bottom right corner and ends in 1023 in the top left corner. The actuators 841 and 197, in red, considered defective, were not used. Therefore, the pupil used is reduced (27 actuators along the diameter of the pupil only) and offseted on the DM.}}
\end{figure}

Out of the 1024 actuators, only 1020 are used because corners are fixed. We number our actuators from 0 (bottom right corner) to 1023 (top left corner) as shown in Figure~\ref{fig:position_pupille_avant}. The four fixed corner actuators are therefore numbers 31, 992 and 1023. The edges of the DM are also composed of fixed actuators, unnumbered. The inter-actuator pitch is $300 \mu$m, for a total size of the DM 9.3 mm. Boston Micromachines announces a subnanometric minimum stroke and a total stroke of 1.5 $\mu$m. All the values presented in this paper, unless stated otherwise, are in mechanical deformation of the DM surface (which are half the phase deformation introduced by a reflection on this surface). The flattened-DM surface quality is valued by Boston Micromachines at 30 nm (root mean square, RMS).

The electronics of the DM allows us to apply voltages between $0$ and $300$ V, coded on 14 bits. The minimum stroke is therefore $300/2^{14}$ V or $18.3$ mV. To protect the surface, the maximum voltage for this DM is $205$ V. 

We use percentage to express the accessible voltages: $0\%$ corresponds to 0 V, while $100\%$ corresponds to a voltage of 205 V. Each percent is thus a voltage of $2.05$ V. The higher the voltage is the more the actuator is pulled towards the DM. A voltage of 100 \% thus provides the minimum value of the stroke, a voltage of 0\% its maximum. The minimum stroke of each actuators is $8.93\,10^{-3}\%$. This value was checked on the THD bench by checking the minimum voltage to produce an effect in the pupil plan after the coronagraph. Gain measurement in Section~\ref{sec:actionneur} will allow us to check the specifications for the maximal and minimum stroke of an actuator. 

Before March 2013, we thought that two actuators were unusable (they did not follow our voltage instructions): the 841 who could follow his neighbors if they were actuated and the 197 that seemed stuck at the 0\% value. 
These actuators will be studied specifically in section~\ref{sec:dead_actioneur}. To avoid these actuators, the pupil before this analysis was reduced (27 actuators across the diameter of the pupil only) and offseted (see Figure~\ref{fig:position_pupille_avant}).

\subsection{Analysis at LAM : interferometric bench and process}

\begin{figure}[ht]
 \begin{center}
  \includegraphics[width = 0.7\textwidth]{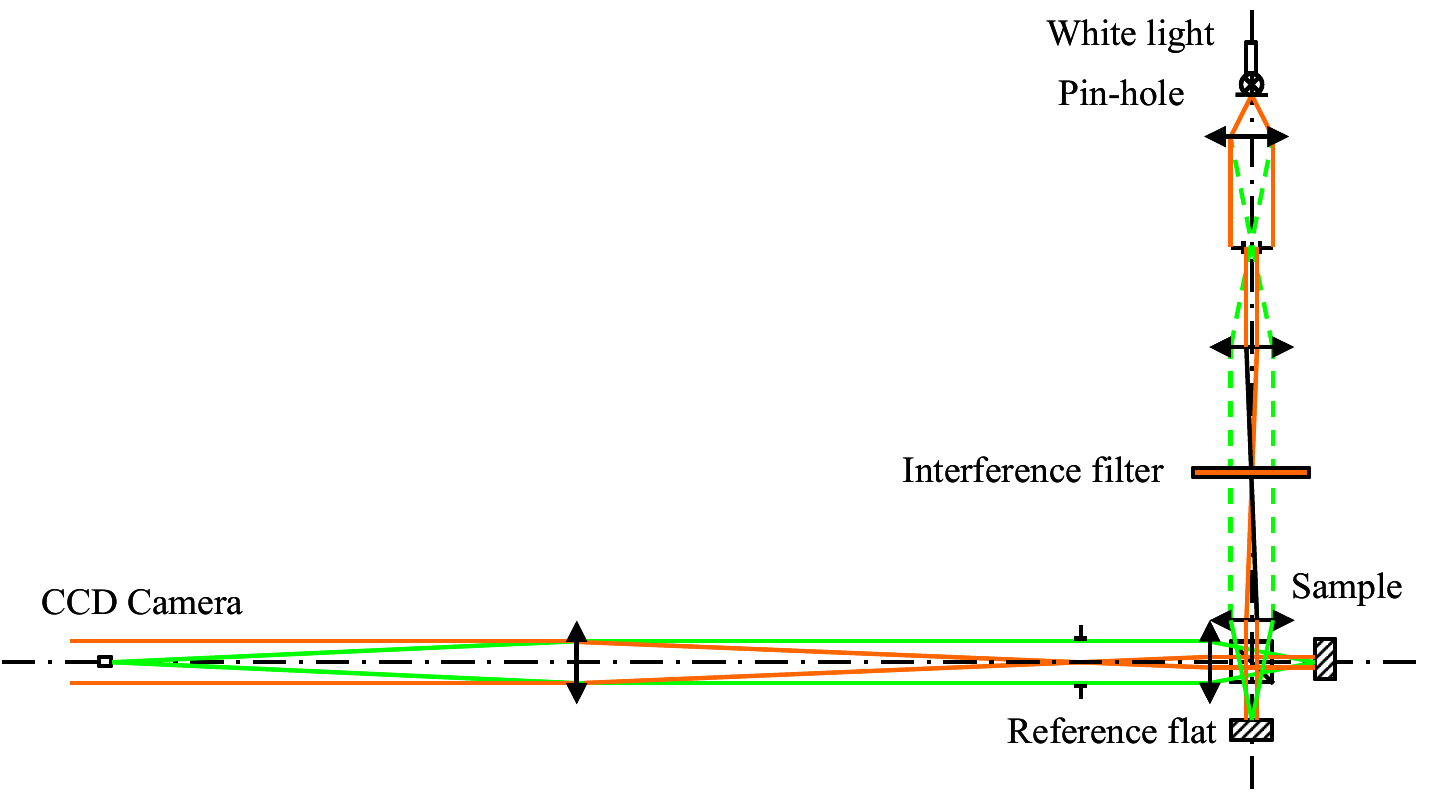}
\end{center}
 \caption[Schéma du banc interférométrique du LAM] 
{ \label{fig:Banc_marseille} The interferometric bench at LAM. Figure from Liotard et al. (2005)\cite{Liotard05}.}
\end{figure}

The interferometric bench at LAM \cite{Liotard05} was developed for the precise analysis of DM. Figure~\ref{fig:Banc_marseille} shows the diagram of the Michelson interferometer. The source is a broadband light, which is filtered spatially by a point hole and spectrally at $\lambda = 650$ nm (this is the wavelength of the THD bench). In the interferometer, one of the mirrors is the DM to analyze (Sample in Figure~\ref{fig:Banc_marseille}). The other is a plane reference mirror (Reference flat in Figure~\ref{fig:Banc_marseille}). At the end of the other arm of the interferometer, a CCD detector (1024x1280) is placed. A lens system can be inserted in front of the camera to choose between a large field (40 mm wide, covering the whole DM) to a smaller field (a little less than 2 mm wide or 6x6 actuators). Both fields will be used in this study. 

The phase measurement is done using the method of Hariharan\cite{Hariharan87}. We introduce 5 phase differences in the reference arm:
\begin{equation}
\label{eq:Hariharan2}
\ \lbrace -2 \pi/2, -\pi/2, 0, \pi/2, 2 \pi/2\rbrace,
\end{equation}
 and record the images with the CCD. The phase difference $\phi$ between the two arms can then be measured using: 
\begin{equation}
\label{eq:Hariharan3}
\ \tan(\phi) = \dfrac{I_{-\pi/2} - I_{\pi/2}}{2 I_0 - I_{-2\pi/2} - I_{2\pi/2}}.
\end{equation}
Assuming a null phase on the reference mirror, the phase on the DM is just $\phi$. Since the phase is only known between 0 and 2$\pi$, the overall phase is unwrapped using a path-following algorithm. This treatment can sometimes be difficult in areas with very high phase gradient. Finally, we measure the surface deformation of the DM by multiplying $\lambda/(2\pi)$ and dividing by 2 (to measure mechanical movement of the DM from optical path difference). The accuracy in the phase measurement is limited by the aberrations of the reference plane mirror and by the differential aberrations in the arms of the interferometer. However, the performance obtained on the measurement of the mechanical deformation of the DM are subnanometric\cite{Liotard05}. We can also retrieve the amplitude on the surface using \cite{Malacara07}: 
\begin{equation}
\label{eq:Hariharan4}
\ \mathcal{M} = \dfrac{3\sqrt{4(I_{-\pi/2} - I_{\pi/2})^2 + (2 I_0 - I_{-2\pi/2} - I_{2\pi/2})^2}}{2(I_{-2\pi/2} + I_{\pi/2} + I_0 + I_{\pi/2} + I_{2\pi/2})}.
\end{equation}
We now present the results of this analysis.

\section{General form of the DM}
\label{sec:forme_miroir}
\begin{figure}[ht]
 \begin{center}
   \includegraphics[trim = 0.8cm 0.2cm 0.8cm 0cm, clip = true, width = 0.495\textwidth]{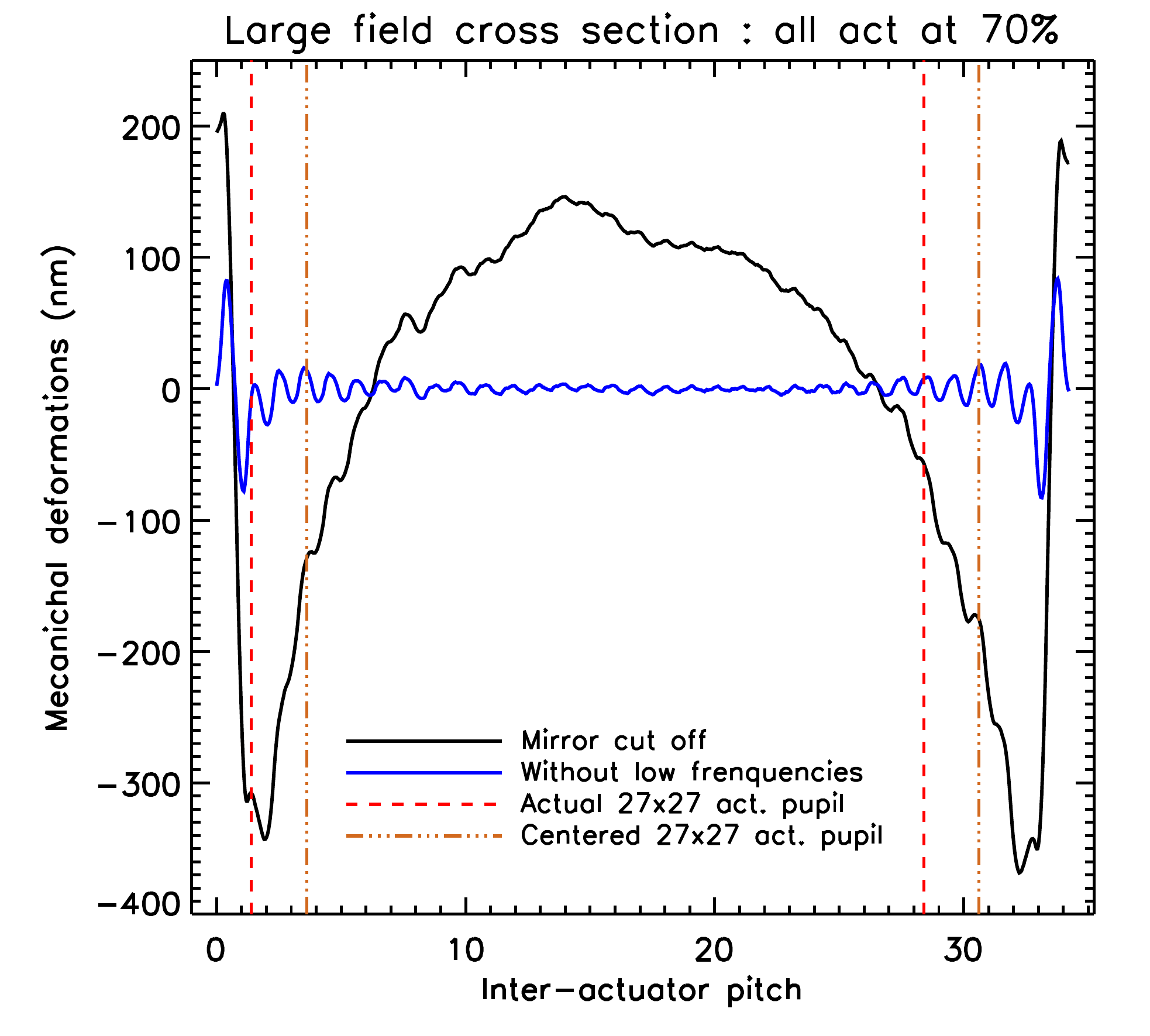}
   \includegraphics[trim = 0.8cm 0.2cm 0.8cm 0cm, clip = true, width = 0.495\textwidth]{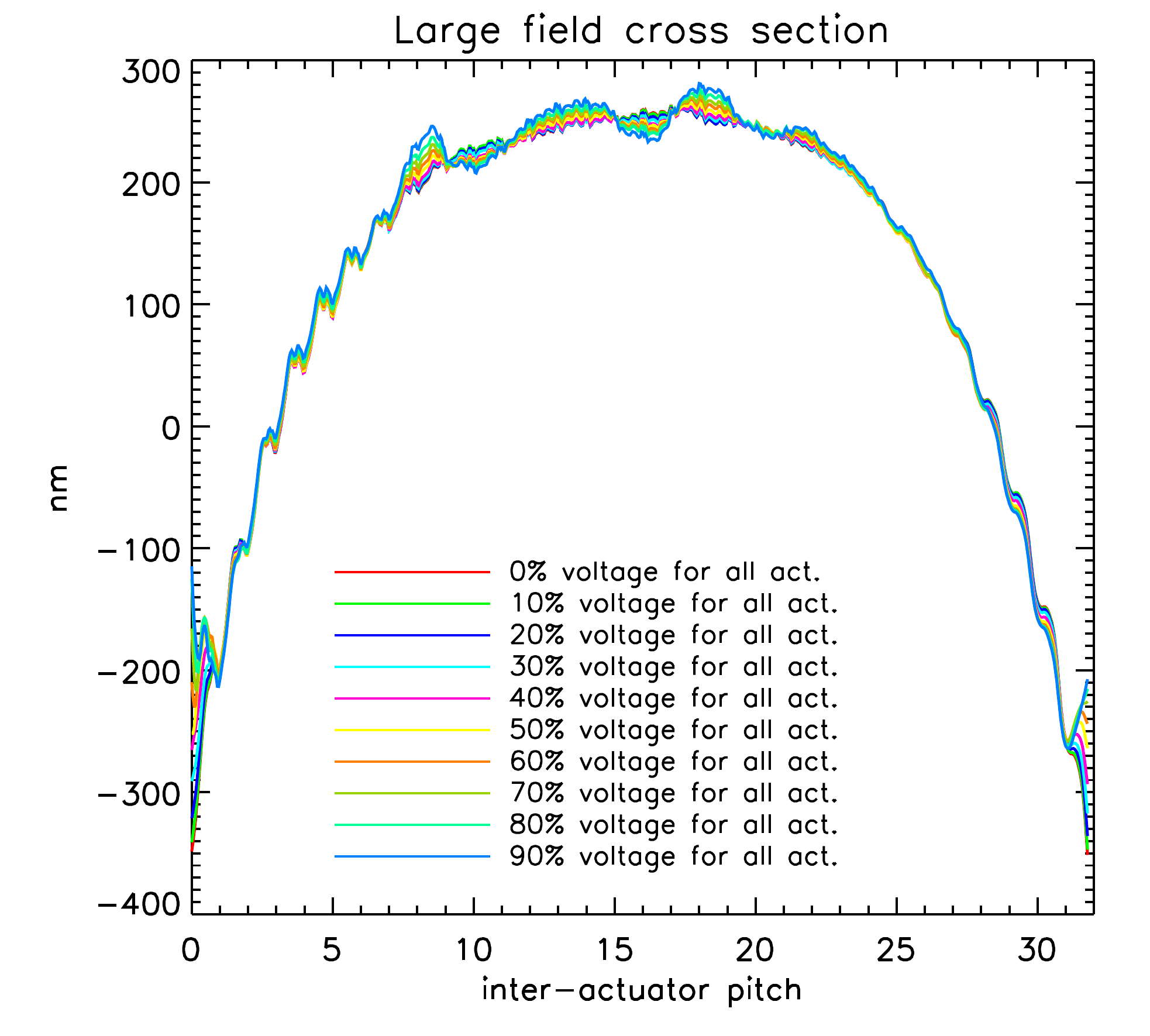}
\end{center}
 \caption[Forme générale du miroir] 
{ \label{fig:coupe_miroir} DM cross sections. Left: cross sections of the whole DM surface in black when all actuators are at 70 \% in voltages. Each point on this curve corresponds to an average over an actuator wide band. In blue line, we plotted the result after removing the frequencies accessible to the DM (in post-treatment, with a smoothing filter). Finally, the vertical lines indicate the limits of the 27x27 actuator pupil before March 2013 (red dotted line) and a pupil centered of the same size (brown dotted line). Right : Same cross sections for different voltages applied to all the actuators (from 0\% to 90 \%). Piston was removed and the curves were superimposed. The abscissas are measured in inter-actuator pitch and the y-axis is in nanometers.}
\end{figure}
Figure~\ref{fig:coupe_miroir} (left) shows, in a black solid curve, a cross section of the DM over the entire surface (in one of the main direction). We applied the same voltage of 70\% to all the actuators. The x-axis is measured in inter-actuator pith and the mechanical deformation in y-axis is in nanometers. The first observation is that a uniform voltage on all the actuators does not correspond to a flat surface on the DM. The general shape is a defocus over the entire surface of approximately $500$ nm (peak-to-valley, PV). The position of the 27x27 actuator pupil on the bench THD before March 2013 is drawn in red vertical lines. The brown vertical lines indicate a pupil of the same size, centered on the DM. The ``natural'' defocus of the DM in a pupil 27 actuators is about 350 nm (PV). 

Figure~\ref{fig:coupe_miroir} (right) represent the same cross section if we apply different uniform tensions to the DM (from 0\% to 90\%). Piston was removed and we superimposed these curves, which show that this form of defocus is present in the same proportions in all voltages. Due to slightly different gains between the actuators, there is a small variations of the actuators at the center between the various applied voltages.

The theoretical stroke of an actuator is $1.5\mu$m and can normally compensates for this defocus by pulling the center actuators of 500 nm while letting the one on the edges at low tensions. However, this correction would be at the cost of a third of this theoretical maximum stroke on the center actuators. The chosen solution on our bench is to place the coronagraph mask outside the focal plane. Indeed, at a distance $d$ the focal plane, the defocus introduced is: 
\begin{equation}
\label{eq:defoc_THD}
      \mathrm{Defoc}_{PV} = \dfrac{d}{8 (F/D)^2},
\end{equation}
in phase difference, in PV, where $F/D$ is the opening of our bench. With the specifcations of our bench, we chose $d = 7$ cm, which correspond to the introduction of a defocus (in phase error) of $700$ nm (PV), which exactly compensates the 350 nm (PV) of defocus (in mechanical stroke) in our 27 actuator pupil. We can therefore chose the voltages around a uniform value on the bench. Before the analysis at the LAM, we chose a voltage of 70 \%, for reasons which are discussed in Section~\ref{sec:actionneur_gain}.

We also note on the black solid line on Figure~\ref{fig:coupe_miroir} (left) the large variation at the edges of the DM (550 nm, PV in only one actuator pitch), when the same voltage of 70 \% is applied to all actuators. This variation tends to decrease when ​​lower voltages are applied to the actuators on the edges. However, this edge must not be included in the pupil. Note that the pupil prior to the analysis in Marseille (in red vertical lines) was very close to these edges.

On the edges, we can clearly seen the ``crenelations'' created by our DM actuators. To measure these deformations, I removed the lower frequencies (including all the frequencies accessible to the DM) numerically with a smoothing filter. The result is plotted in Figure~\ref{fig:coupe_miroir} (left) in blue solid line. We clearly see this crenelation effect increase as we approache the edges. Once again, it is better to center the pupil on the detector to avoid the edge actuators. These effects are the main causes of the poor surface quality of MEMS DMs that we discussed in the next section.

\section{Surface quality}
\label{sec:qualite_surface_DM}

\begin{figure}[ht!]
 \begin{center}
  \includegraphics[width = 0.495\textwidth]{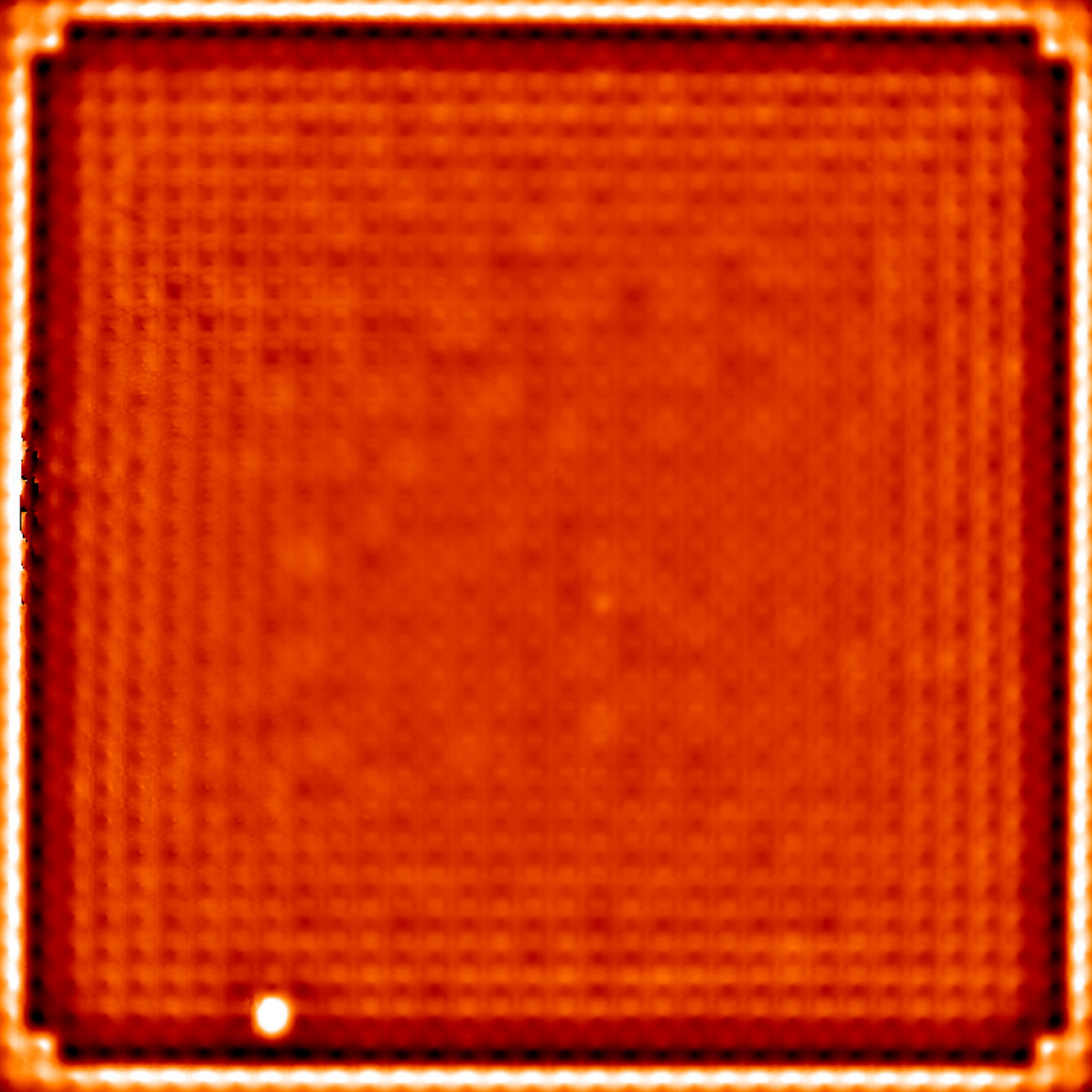}
  \includegraphics[width = 0.495\textwidth]{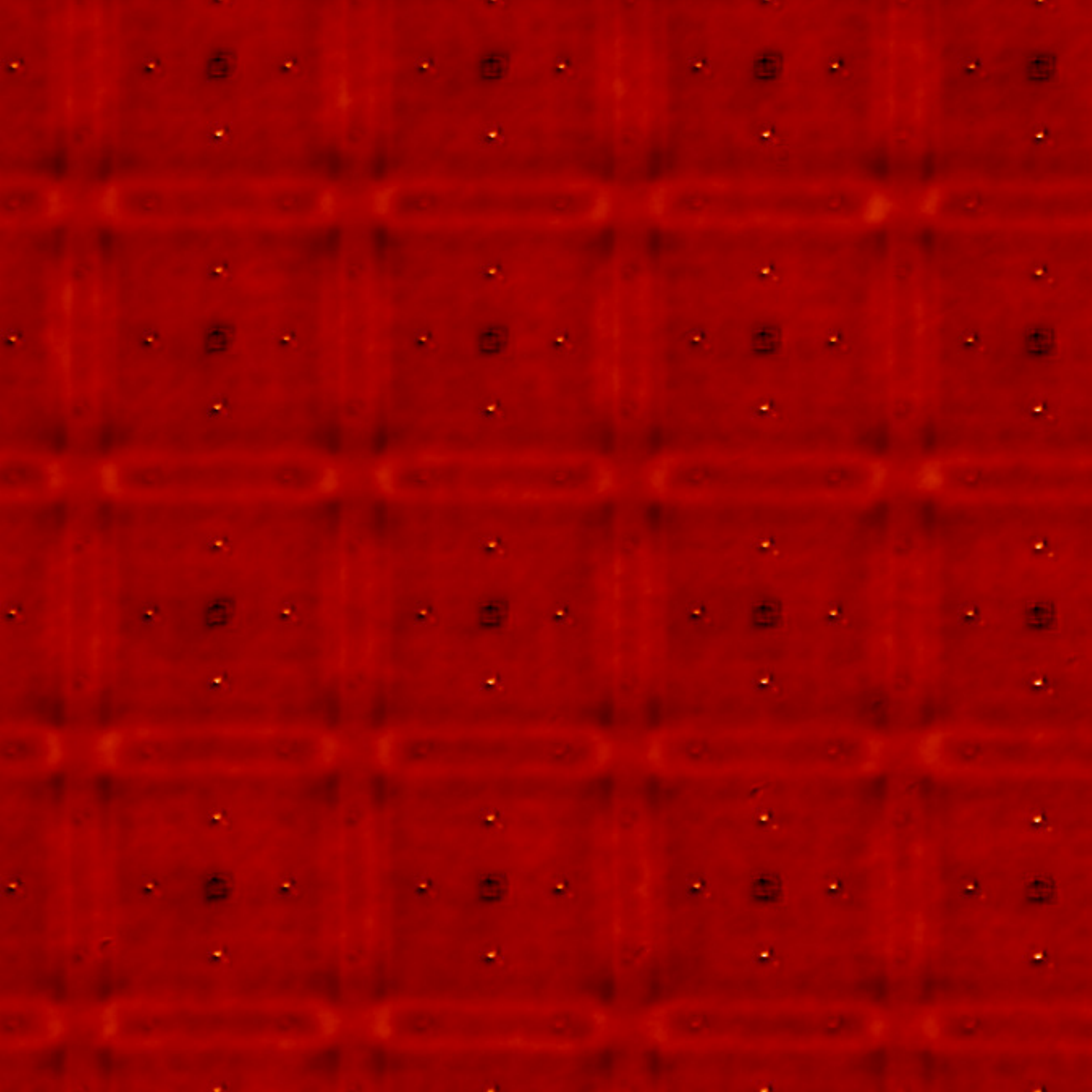}
\end{center}
 \caption[Surface du miroir et des actionneurs] 
{ \label{fig:surfaces_DM} Surface of the DM in large field on the left (the whole DM is about 10 mm by 10 mm) and in small field  on the right, centered on the 4 by 4 central actuators (ie 1.2 mm by 1.2 mm). In both cases, all the actuators are set to 70 \% in voltages. To observe the fine structures of the DM, we removed in both cases low frequencies digitally in post-processing. On the left image, we can see actuator 769 (bottom left), which is fixed to 0\% (see Section~\ref{sec:Act_mort_mort}).}
\end{figure}

We now study the surface of our DM first through the level of aberrations, then with the study of the Power Spectral Density (PSD). 

In Fig~\ref{fig:surfaces_DM} we show images of the surface of the DM obtained in large field (about 10 mm by 10 mm) on the left and small field (right), centered on the 4 by 4 central actuators (ie 1.2 mm by 1.2 mm). In both case, we removed all frequencies reachable by the DM (below 0.5 (inter-actuator pitch) $^{-1}$) through a smoothing filter in post processing to observe its fine structures. For example, the defocus mentioned in the previous section has been removed. In the large field, we can observe the actuator 769 (in the lower left), which is fixed to the value 0\% (see Section~\ref{sec:Act_mort_mort}) and was unnoticed before this analysis but no noticeable sign of the two known faulty actuators (see Figure~\ref{fig:position_pupille_avant}). We also note the edges and corners, very bright, due to fixed actuators. 

Boston Micromachines announces a surface quality 30 nm (RMS) on the whole DM when it is in a ``flat'' position. Because of the high defocus defect that we corrected using a defocus of the coronagraphic mask, we have not tried to obtain a flat surface on the DM to verify this number. However, an estimate of the remaining aberrations in a ``flat'' position can be maid by removing in post processsing all the frequencies correctable by our DM. We measure the remaining aberrations without the edges and found 32 nm (RMS). This is slightly higher than the specifications of Boston Micromachines but  one of the actuator at least is broken. The same measurement on our actual 27 actuators offseted pupil gives 8 nm (RMS) and 7 nm (RMS) for a same size pupil centered.

In the right image, we observe the details of the actuator. We observed three types of deformations:
\begin{itemize}
\item the center of the actuator, in black, with a size of about 25 $\mu$m
\item the edges, which appear as two parallel lines separated by $45 \mu$m and of length one inter-actuator pitch (300$\mu$m)
\item the release-etch holes of the membrane (4 in the central surface + 2 between the parallel lines of the edges). In the principal direction of the DM, they appears very 150 $\mu$m and are only a few $\mu$m larges. \textit{A priori}, these holes are a consequence of the making process by lithography.
\end{itemize}

\begin{figure}[]
 \begin{center}
   \includegraphics[width = 0.495\textwidth]{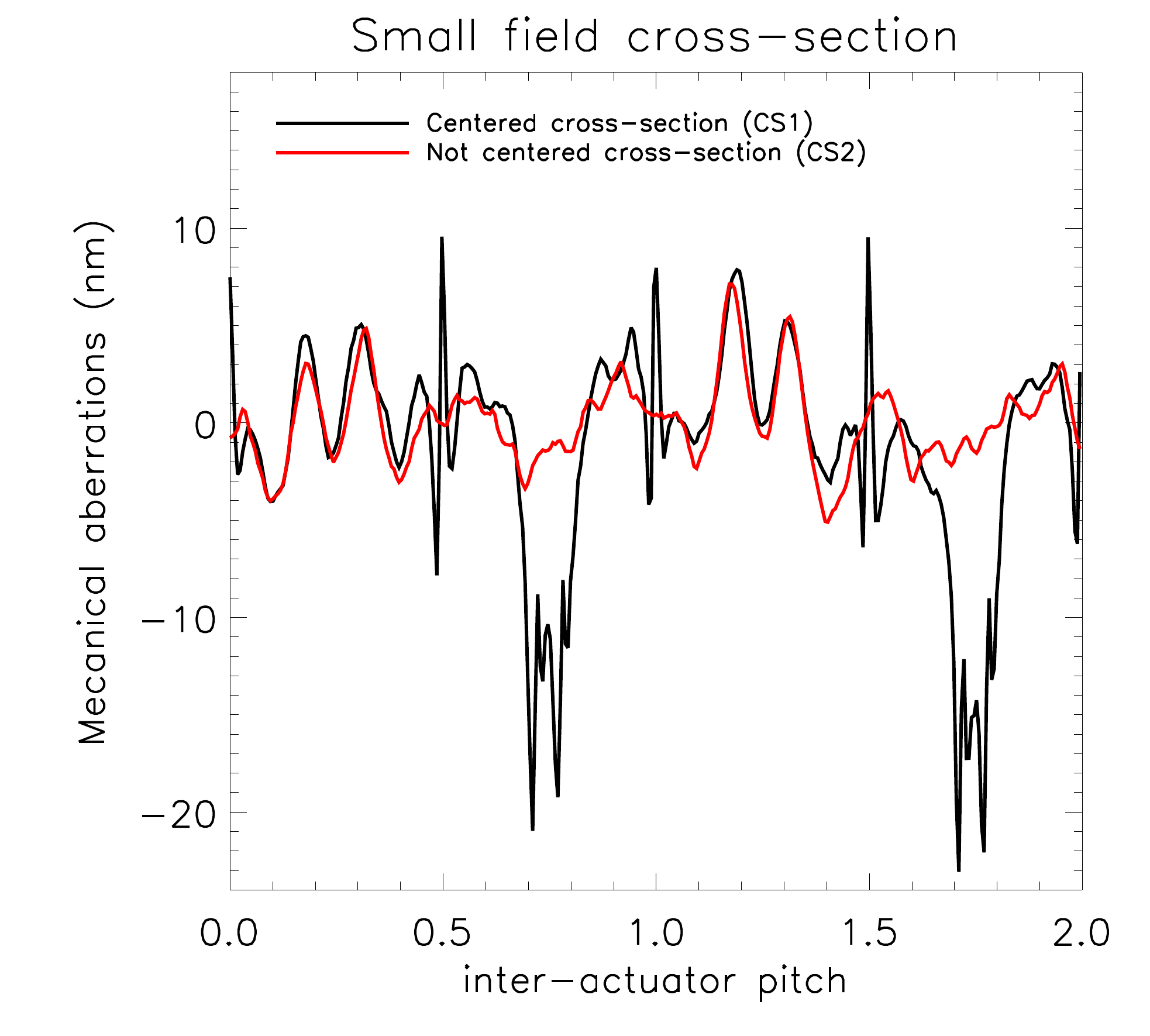}
  \includegraphics[width = 0.495\textwidth]{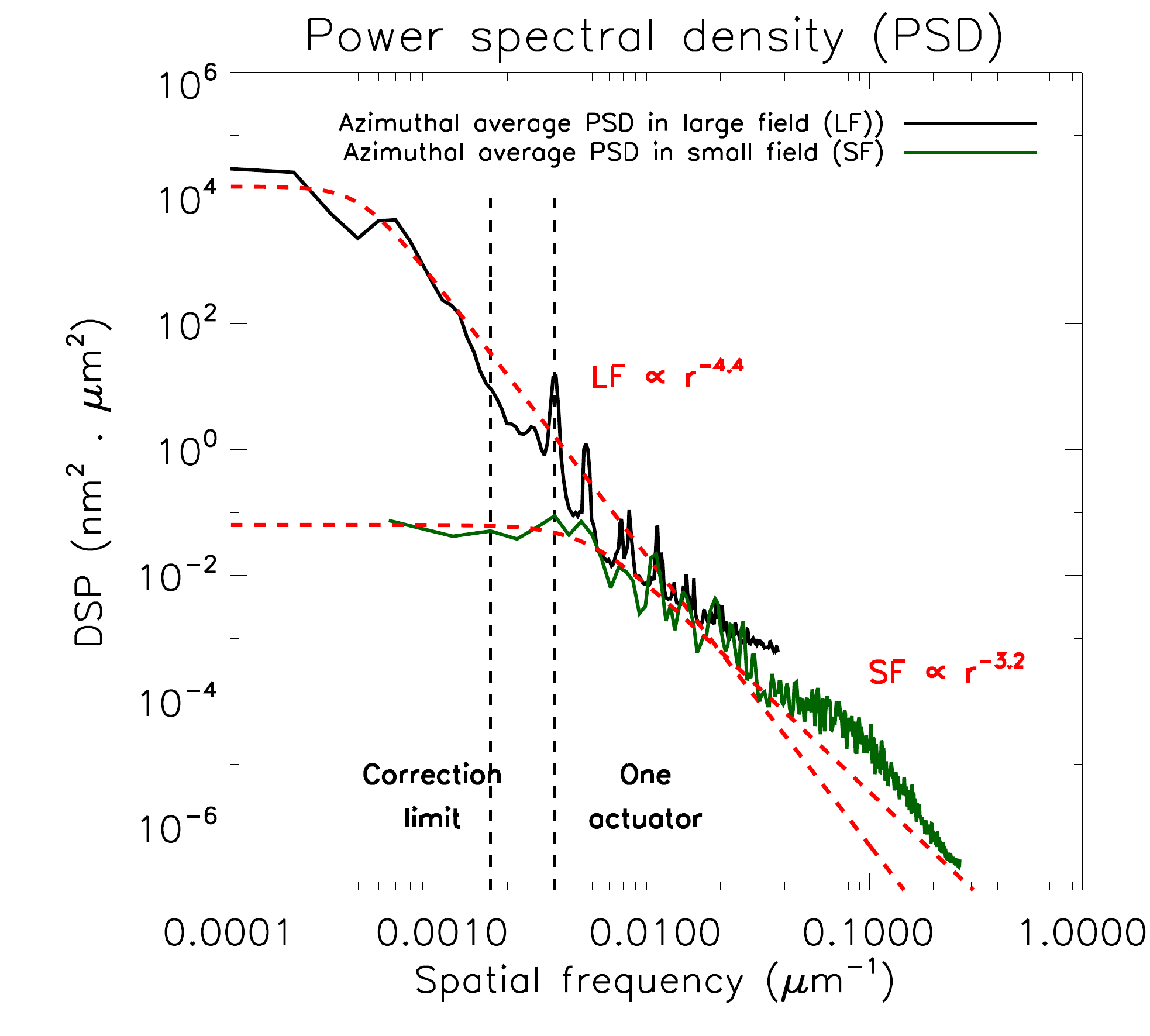}
\end{center}
 \caption[coupe_DSP] 
{ \label{fig:coupes_act_dsp_DM} On the left, cross sections on two actuators observed in small field. Each point of these cross sections is an average over a width of 0.1 inter-actuator pitch, either avoiding center and release-etch holes (curve ``best case'', in red) or on the contrary right in the center of an actuator (curve `` worst case'' in black). On the right, azimuthally averaged PSD measured on the whole DM (wide field) and on some actuators (narrow field). The frequencies on the horizontal axis are measured in $\mu$m$^{-1}$ and the vertical axis is in nm$^2 . \mu$m$^{2}$. The black dotted vertical lines indicate remarkable frequencies: the frequency of the actuators ($1/300\mu$m$^{-1}$) and the maximum correctable frequency by the DM, of (2 inter-actuator pitch)$^{-1}$, or $1/(2*300) \mu$m$^{-1}$. Finally, in red, we adjusted asymptotic curves.}
\end{figure}

We measured cross sections along two actuators in the small field, shown in Figure~\ref{fig:coupes_act_dsp_DM}. The horizontal axis is in inter-actuator pitch and the vertical axis is in nanometers. Each of the points of these cross-sections is an average on a width of about 0.1 inter-actuator pitch. We placed these bands either right in the center of an actuator (curve ``worst case'', in black) or in a way to avoid both centers and release-etch holes (curve ``best case'', in red). The two bumps in 0.15 and 0.35 and in 1.15 and 1.35 inter-actuator pitch, common to both curves correspond to the parallel lines at the edges of the actuators. They produce mechanical aberrations of 12 nm (PV). The centers, in the black curve, are located in 0.65 and 1.65 inter-actuator pitch. They introduce mechanical aberrations 25 nm (PV). It is not certain that the aberrations in the release-etch holes are properly retrieved for several reasons. First, their size is a smaller than 0.1 inter-actuator pitch, so they are averaged in the cross section. We are also not sure that the phase is correctly retrieved in the unwrapping process as it encounters a strong phase gradient in these holes. They produce aberrations of 20 nm (PV). In total, on one actuator, aberrations of 30 nm (PV) and 6 nm (RMS) are obtained. 

Figure~\ref{fig:coupes_act_dsp_DM} shows the azimuthally averaged DSP of the DM. The black curve represents the azimuthally averaged DSP for the whole DM (large field). We clearly observe the peak at the characteristic frequency of the DM ($1/300\,\mu$m$^{-1}$), indicated by a black dotted line. We can see peaks at other characteristic frequencies ($1/(300*\sqrt{2})\,\mu$m$^{-1}$, $2/(300)\,\mu$m$^{-1}$, ...). We took an azimuthal average to average these frequencies and observe a general trend. We repeated the same operation for DSP calculated on a small field. As shown in red in Figure \ref{fig:coupes_act_dsp_DM} (right), we plotted the trends of these azimuthal DSP, which shows a asymptotic behavior in $-4.4$ for the big field and $-3.3$ for small field. Indeed, very small defects can come from differential aberrations in the interferometer, deformation of the flat reference mirror or noise in the measurement.  We therefore adopt the large field value of f$^{-4.4}$ for asymptotic behavior.

We now precisely study the behavior of a single actuator (Section~\ref{sec:actionneur}), ie the influence function, the coupling with its neighbors, and the gain when we applied different voltages. 
 
\section{Behavior of a single actuator}
\label{sec:actionneur}
\begin{figure}[ht]
 \parbox{0.26\textwidth}{ \centering
 \includegraphics[width = 0.25\textwidth]{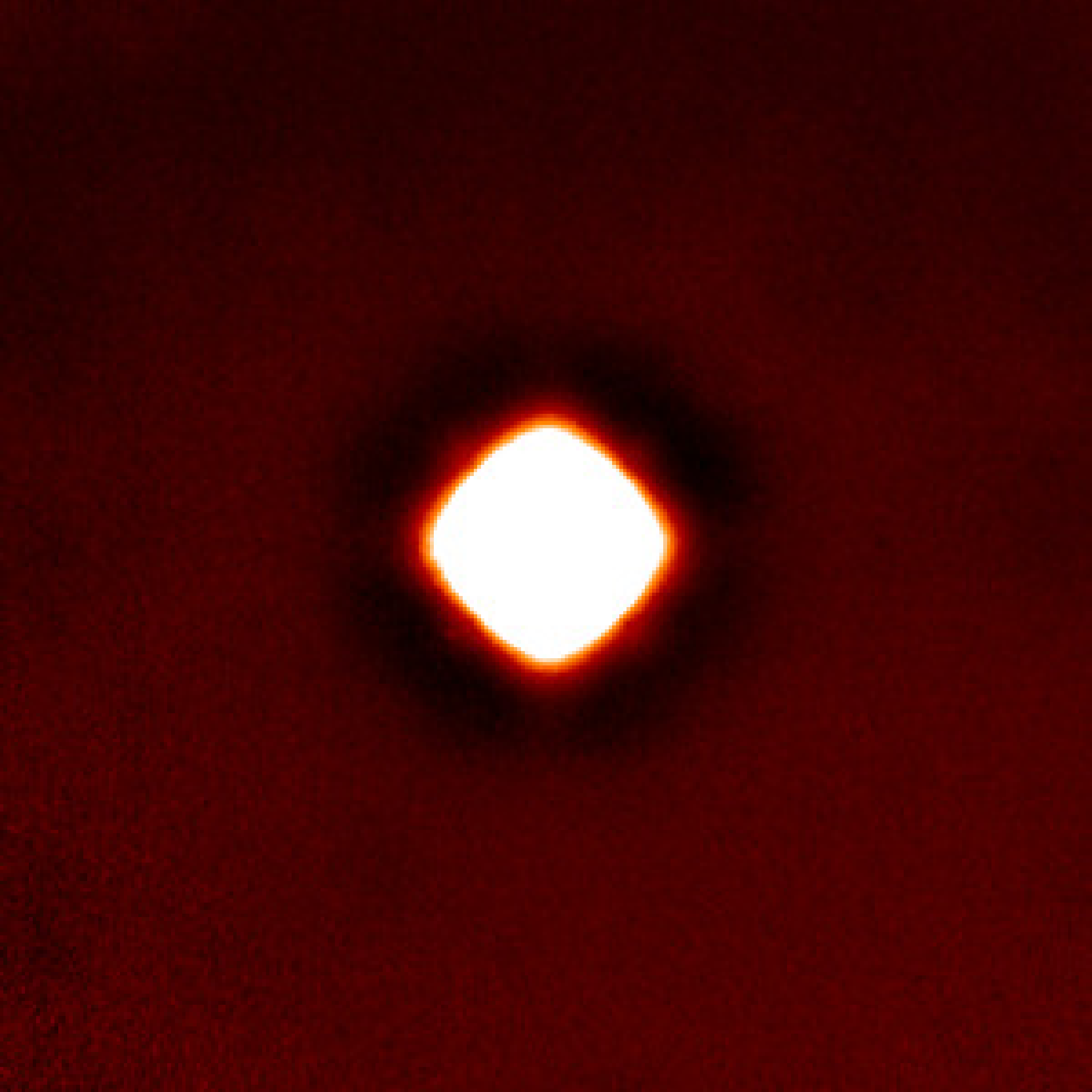}}\parbox{0.74\textwidth}{\centering \includegraphics[trim = 0.8cm 0.2cm 0.8cm 0cm, clip = true, width = 0.36\textwidth]{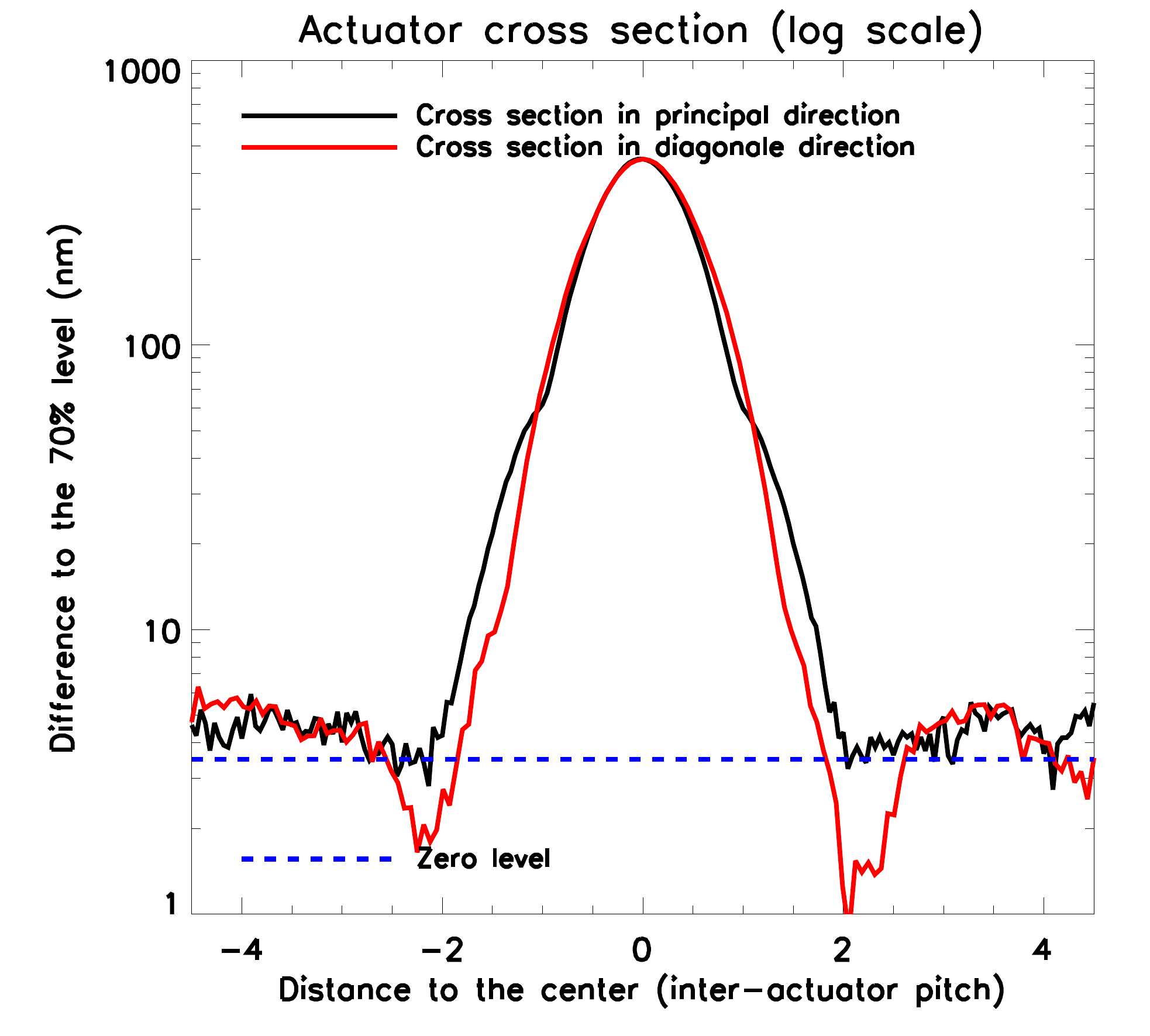}
 \includegraphics[trim = 0.8cm 0.2cm 0.8cm 0cm, clip = true, width = 0.36\textwidth]{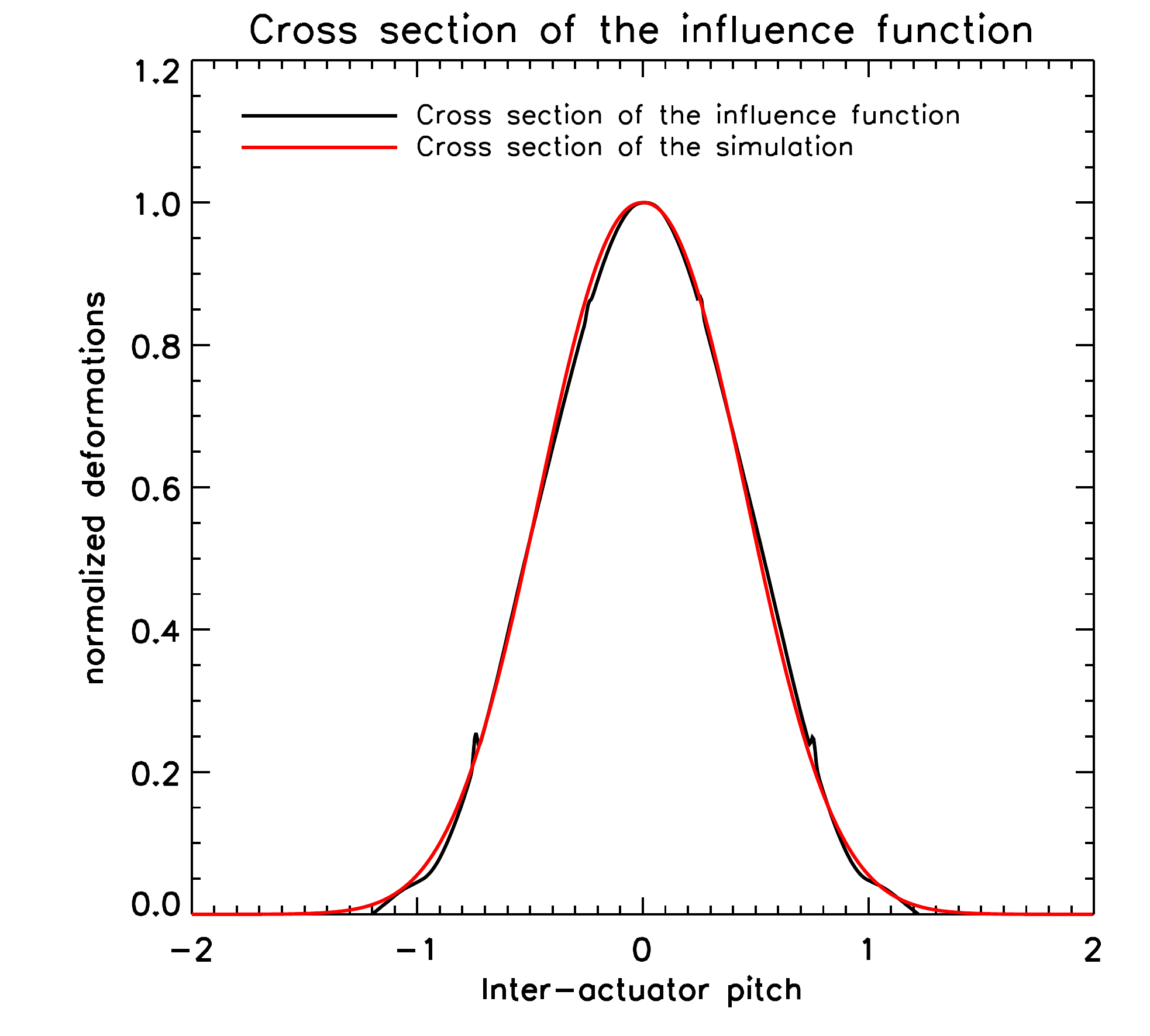}
 }
\caption[influencefonction] 
{ \label{fig:etude528_infl} Influence function. Left, measurement of the influence function of a central actuator. Center, cross section of the influence function in logarithmic scale along a principal direction of the mirror and in an diagonal direction. Right, cross section of the influence function along a principal direction, on which is superimposed  cross section of a simulated influence function. The abscissas are in inter-actuator pitch and the vertical axis are in nanometers.}      
\end{figure}
For this analysis, we observe the behavior of a central actuator (number 528). We will measure its influence function and the inter-actuator coupling then study its gain, maximum and minimum strokes. These measurements were conducted by applying to the actuator 528 several voltages ranging from 10 to 90 \% while the rest of the actuators are set at the value 70 \%.

\subsection{Influence function and coupling}
\label{sec:actionneur_fonction_influence}

We study the influence function $IF$ of an actuator, movement of the surface when a voltage is applied. This influence function can be simulate using \cite{Huang08}: 
\begin{equation}
\label{eq:Huang08}
\ IF(\rho) =  \exp[\ln(\omega) (\dfrac{\rho}{d_0})^\alpha],
\end{equation}
where $\omega$ is the inter-actuator coupling and $d_0$ is the inter-actuator pitch.

Figure~\ref{fig:etude528_infl} shows the influence function of a central actuator (528). At first, we apply a voltage of 40 \% to the actuator (the others remaining at a voltage of 70 \%) then a voltage of 70 \% and made ​​a difference, shown in the left picture. This is therefore the influence function for a voltage of $-30 \%$. We can observed that the influence function has no rotational symmetry. The main shape is a square, surrounded by a small negative halo.

We made cross section in several directions: one of the main directions of the DM, one of the diagonals. The results are presented on a logarithmic scale in Figure~\ref{fig:etude528_infl} (center). The distance to the center of the actuator is in inter-actuator pitch. We applied an offset to plot negative values in a logarithmic scale and we indicate the zero level by a dotted blue line. In the main direction, a break in the slope is observed at a distance of 1 inter-actuator pitch. The influence of the actuator in this direction is limited to 2 inter-actuator pitch in each way. In the diagonal direction, the secondary halo is about 3 nm deep, which is 0.5 \% of the maximum. Due to this halo, the influence is somewhat greater (however, less than 3 inter-actuator pitch). 

On Figure~\ref{fig:etude528_infl} (right), is plotted a cross section of the influence function in a principal direction of the DM. The inter-actuator coupling (height of the function at the distance of 1 inter-actuator pitch) is of 12 \%.  We fitted a curve using the function described in Equation~\ref{eq:Huang08} using this coupling and found $\alpha = 1.9$. This shows that the central part of the influence function is almost a Gaussian ($ \ alpha = $ 2), but do not take into account the ``wings''.

\subsection{Gain study}
\label{sec:actionneur_gain}
\begin{figure}[]
 \begin{center}
  \parbox{0.33\textwidth}{ \centering
  \includegraphics[height = 5.55cm]{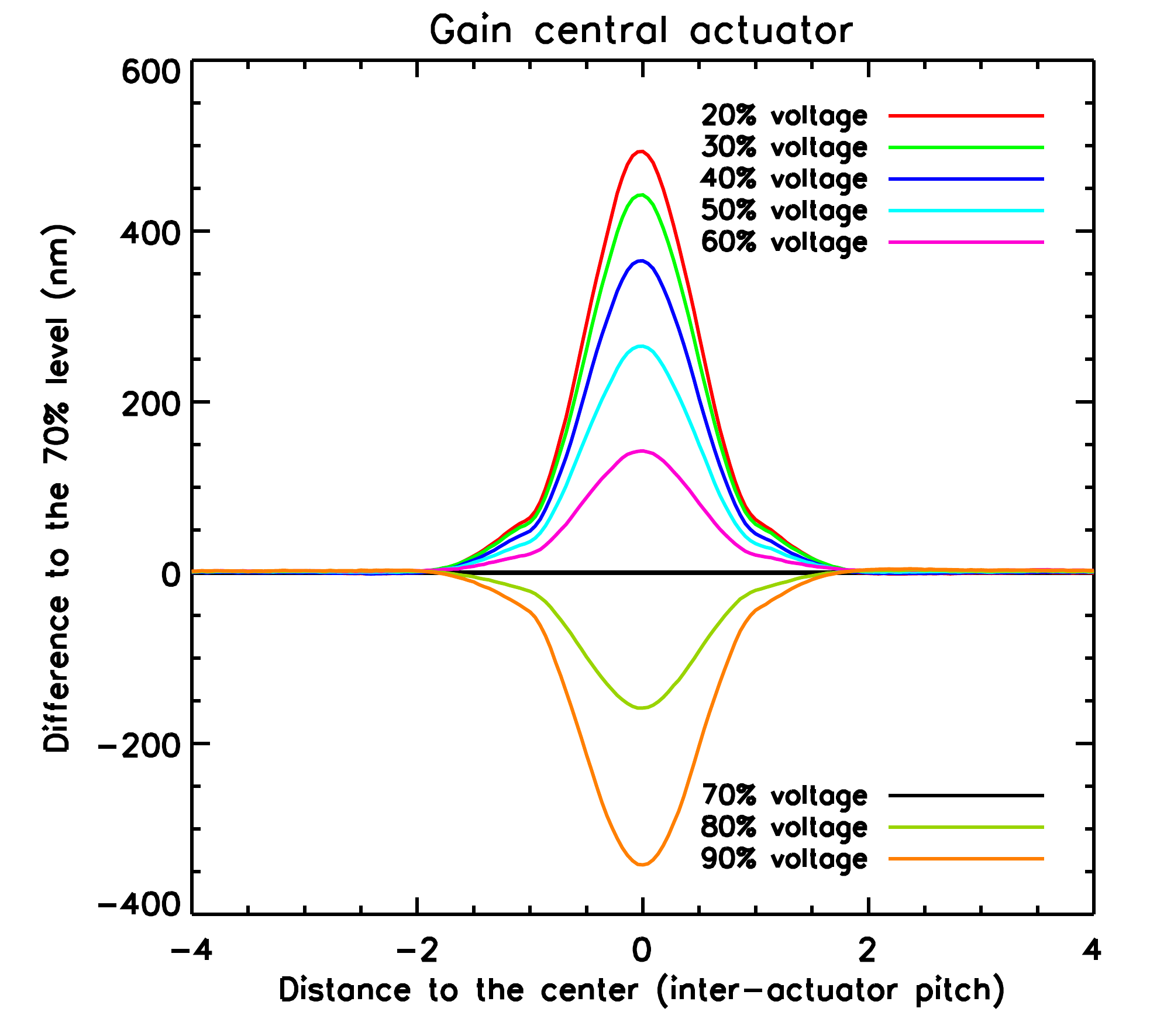}}
  \parbox{0.66\textwidth}{ \centering
  \includegraphics[height = 6cm]{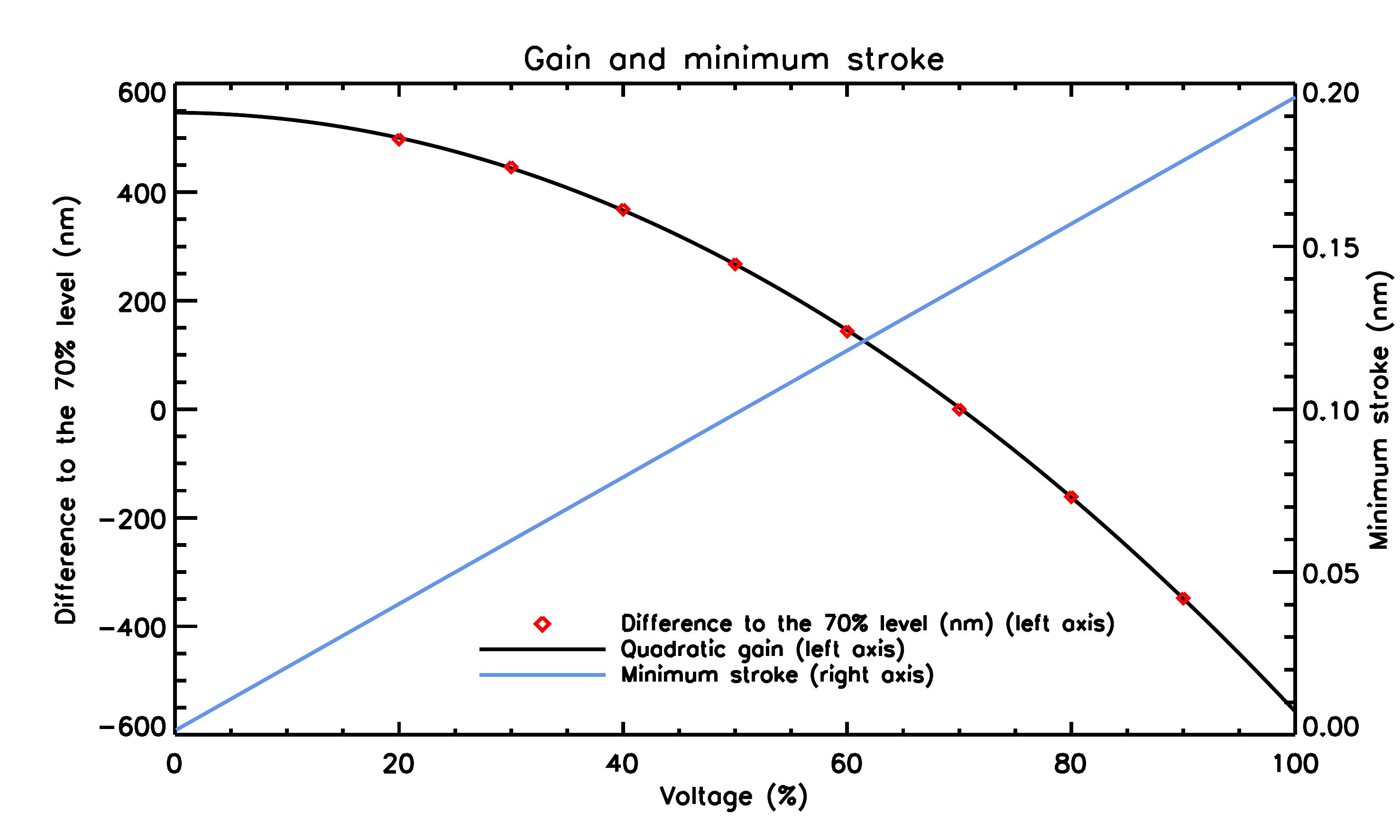}} 
\end{center}
 \caption[course_gain] 
{ \label{fig:etude528_gains} Study of one actuator: stroke and gain. Left: influence functions for different applied voltages. Right: maximum values ​​of these influence functions in red and quadratic gain (black solid curve). The minimum percentage applicable ($8.93\,10^{-3}$\%) can produce different minimum stroke depending on your position on this quadratic curve : we plot the minimum stroke around each voltage in blue (the scale of this curve, in nanometers, can be read on the right axis).}
\end{figure}
In this section, we measured the maximum of the influence function for different voltages applied to the 528 actuator, the other remaining 70 \%. Figure~\ref{fig:etude528_gains} (left) shows the superposition of cross sections in a principal direction of the DM for voltage values of ​​20 \%, 30 \%, 40 \%, 50 \%, 60 \%, 70 \%, 80 \%, 90 \%. We fitted Gaussian curves for these functions and observed that the maximum values of the peak are always located at the same place, and the width of the Gaussian is constant for the range of applied voltages. This shows that the influence function is identical for all the applied voltages. We plot the maximum of these curves as a function of the voltages in red diamonds in Figure~\ref{fig:etude528_gains} (right). The scale of these maximum can be read on the left axis, in nanometers. We then adjusted a quadratic gain (black solid curve) on this figure. This allows us to extrapolate the voltage values for ​​0 \% and 100 \%. From this figure, it can be deduced that:

\begin{itemize}
\item the maximum stroke is 1100 nm ($1.1 \mu$)m slightly less than the value indicated by Boston Micromachines.
\item the value of 70 \% is the one that allows the maximum stroke in both ways (545 nm when we push and 560 nm when we pull). If the actuator is at a value of 25 \%, we can only enjoy a maximum stroke of 140 nm in one direction. For this reason, we used to use the DM at values around 70 \% before March 2013.
\item the gain have a quadratic variation and therefore, the value of the minimum stroke in volt or in percent ($8.93 \, 10^{-3}\%$) corresponds to different minimum strokes in nanometers depending on the location on this curve. We plotted the value of the minimum stroke in blue on the same plot (the scale of this curve, in nanometers, can be read on the right axis). We observe that a variation of $ 8.93 \,10^{-3}$\% around 70 \% produces a minimum stroke of $0.14$ nm, which is twice the movement produced by the same variation around 25 \% ($0.07$ nm).
\end{itemize}

Applying voltages around 70 \% makes sense if we try to make the most of the stroke of the DM, but if we try to correct for small phase aberrations (which is our use of this DM), we should apply the lowest voltages possible.

We observed the positions of all the actuators and verified that they are evenly distributed on the surface. The gains of all actuators are very close on all the surface (variation of 20\% between the minimum gain and the maximum). 

We finish this study by an inventory of the different failures that we encountered and the solutions that we have fortunately been able to put in place to overcome these failures.

\section{Damaged actuators}
\label{sec:dead_actioneur}

Before the analysis in March 2013, the actuators 841 and 197 were not responding correctly to our commands. A specific study on these actuators allowed us to overcome these dysfunctions and include them again in the pupil. 

\subsection{The slow actuator}
\label{sec:act_mou}
\begin{figure}[]
 \begin{center}
  \includegraphics[width = 0.4\textwidth]{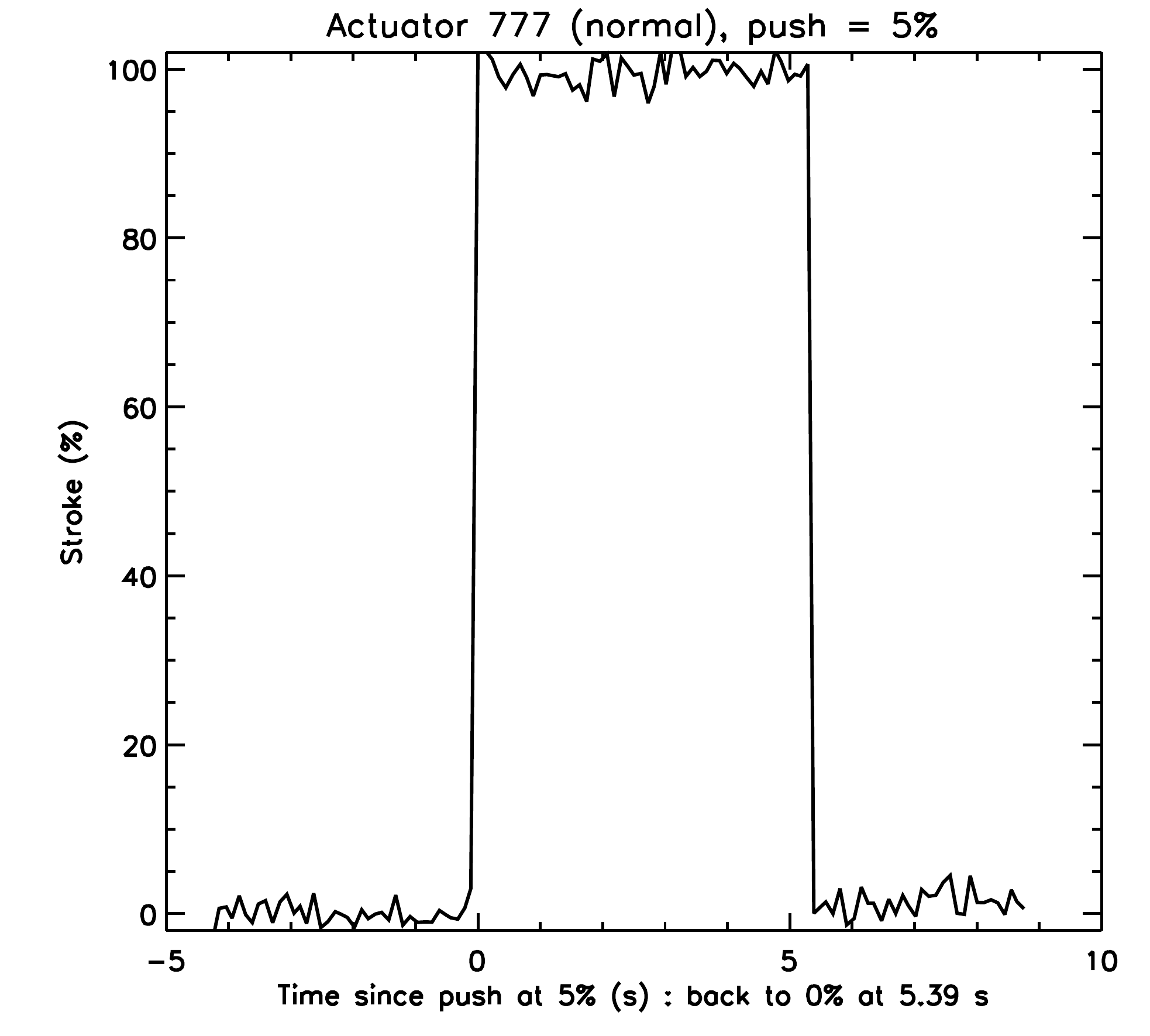}
  \includegraphics[width = 0.4\textwidth]{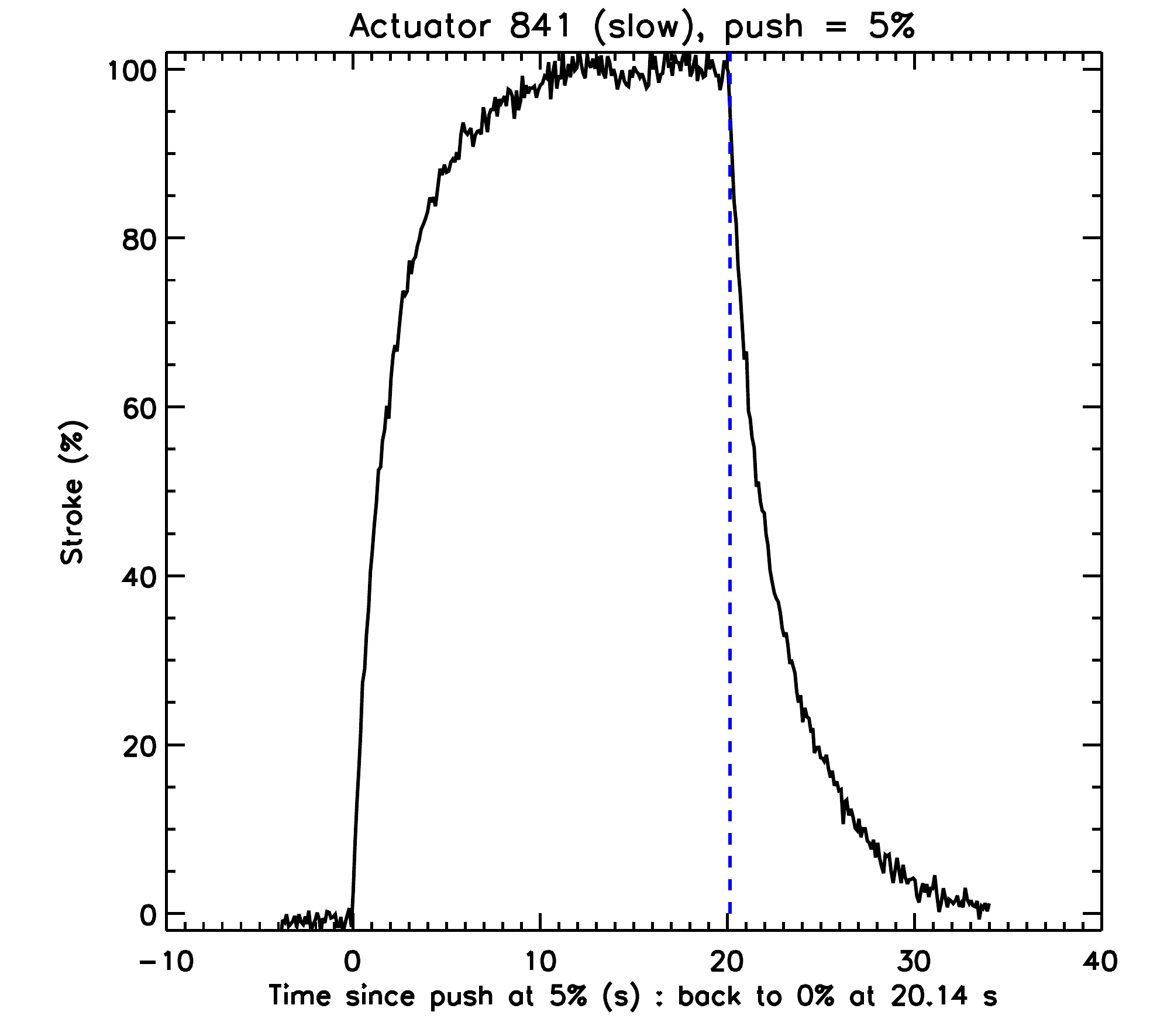}
\end{center}
 \caption[slow_act] 
{ \label{fig:act_mou} Study of the slow actuator. Temporal response to a $+5\%$ command in voltage for a normal actuator (left) and for the actuator 841 (right). Starting with a voltage of $70\%$, we send a $+5\%$ at 0s, wait for this command to be applied and then send a  $-5\% $ command, at 5.39 s for normal actuator and at 20.14 s for the actuator 841. The vertical axis is $\%$ of the stroke, the abscissa in seconds since the update command of $+5\%$.
The dashed blue indicates the sending of the $-5\%$ command.}
\end{figure}
We found that the actuator 841 responded to our  voltage commands but with a very long response time. The interferometric bench in Marseille is not suited for temporal study (the successive path differences introductions limit the measurement frequency). Therefore, I used the phase measurement method developed on the THD bench: the self-coherent camera, see Mazoyer et al. (2013)\cite{Mazoyer13}. I examined the temporal response of the 841 actuator after a command of $+ 5 \% $ and compared it with the temporal response of a normal actuator (777). From a starting level of $70\%$ for all of the actuator of the DM, we first sent a command to go to $75\%$ to each of these two actuators, wait for this command to be executed and sent an order to return to the initial voltage. Figure~\ref{fig:act_mou} shows the results of this operation for a normal actuator (777, left) and for the slow actuator (841, right). The measurement frequency is on average 105 ms. The horizontal axis is the time (in seconds), with origin the date at which the $+5\%$ command is sent. Our phase measurement method does not give an absolute measurement of the phase and so we normalized the result (0\% is the mean level before the command, 100\% is the mean level after the $+5\%$command.

For the normal actuator, the response time is inferior to the measurement period (105 ms in average). This result is consistent with the response time of an actuator announced by Boston Micromachines ($< 20\mu$s) although we cannot verify this value with this method. For the slow actuator, there is a much slower response to the rise as well as to the descent. We measured the response time to $95\% $ of the maximum in the rise (7.5 s) and in the descent (8.1 s). However, as the static gain of the actuator is comparable with the gain of healthy actuators, we deduce that this actuator goes slowly but surely to the right position. 

\begin{figure}[ht!]
 \parbox{0.65\textwidth}{ \centering
  \includegraphics[trim = 0.5cm 2.5cm 0.52cm 2.5cm, clip = true, width = 0.645\textwidth]{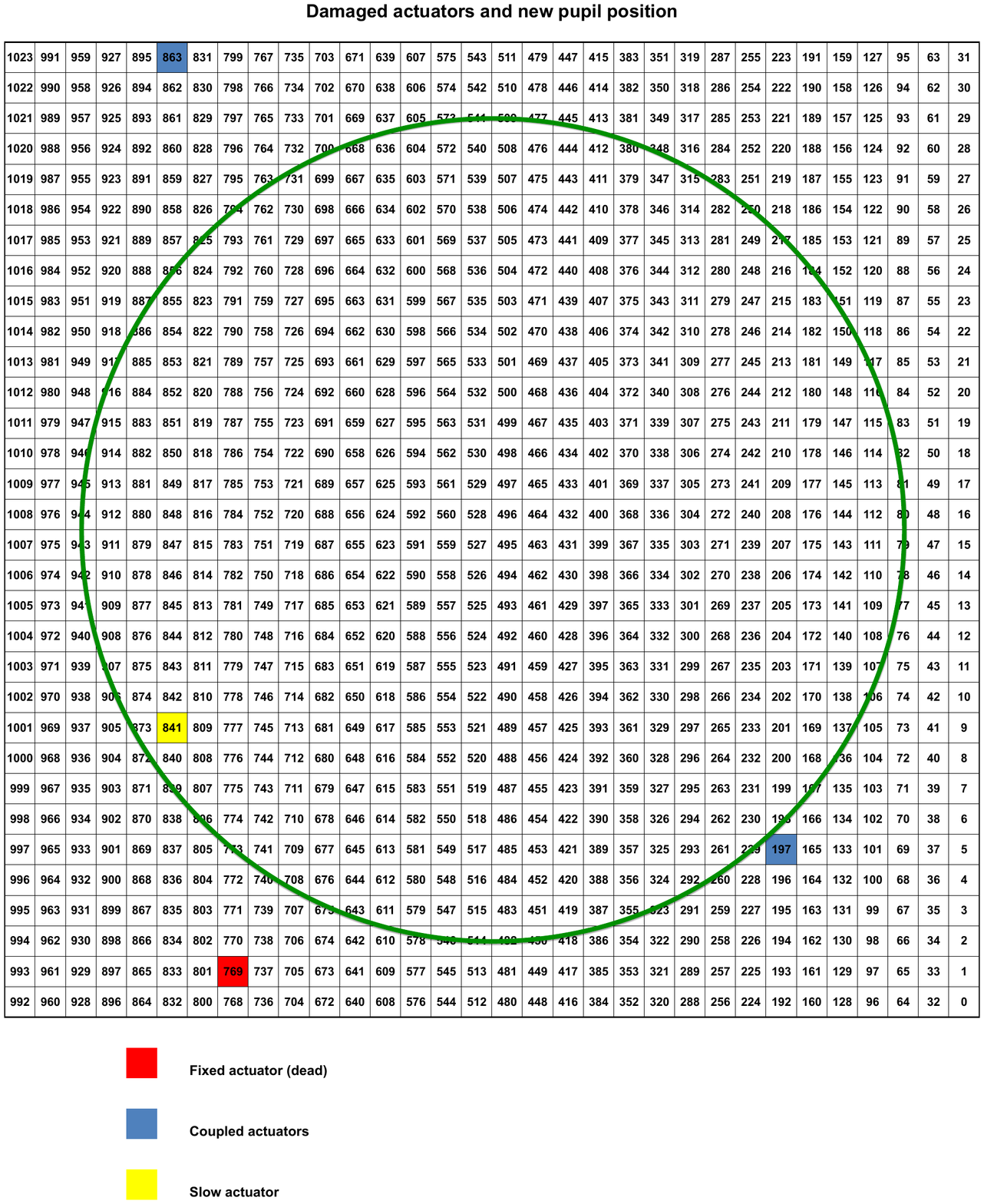}}
\parbox{0.34\textwidth}{\centering \caption[pupapres] 
{ \label{fig:position_pupille_apres} This study allowed us to identify precisely the causes of actuator failures and recenter the pupil on the DM, including actuators 197 and 841.}}    
\end{figure}

\subsection{The coupled actuators}
\label{sec:Act_couples}

We realized that the actuator number 197 responded to the commands applied to the actuator 863, at the other end of the DM. It seems that the actuator 197 has a certain autonomy, but in case of large voltage differences applied on these two actuators, the 197 follows the commands applied to the 863 actuator. We carefully verified that if we apply the same voltage to them, these two actuators respond correctly to the command and have comparable gains than the other actuators. The actuator 863 is fortunately on the edge of the DM, so we can center the pupil with no influence of this actuator. Since, we have recentered the pupil to include the the 197 actuator back (see Figure~\ref{fig:position_pupille_apres}). We systematically apply same voltages to both actuators simultaneously.

\subsection{The dead actuator}
\label{sec:Act_mort_mort}

Finally, we notice that the 769 actuator does not respond at all to our commands. This actuator is on the far edge of the DM. It is possible that he broke during the transportation towards LAM laboratory, but as it was far off pupil, we may have previously missed this failure. This actuator is fixed to the value 0\% regardless of the applied voltage. However, we checked that it has no influence over 2 inter-actuator pitch.

\section{Conclusion and consequences on the bench}
\label{sec:consequences_DM}

The identification of the faulty actuators and the solutions to overcome these dysfunctions have enabled us to recenter the pupil on our DM. Figure~\ref{fig:position_pupille_apres} shows the position of the pupil on the DM after the study at LAM.  This centering has enabled to move away from the edges of the DM. We also saw that this centering is preferable to limit the introduction into the pupil of aberrations at high frequencies non reachable by the DM. Finally, we recently lowered the average value of the voltages on the DM from 70 \% to 25 \% and improve by a factor of 2 the minimum stroke reachable by each actuators. These upgrades played an important role in the improvement of our performance on the THD bench. To see the latest results on this high contrast bench, see Galicher et al. (2014)\cite{Galicher14} and Delorme et al. (2014)\cite{Delorme14}.

%%%%%%%%%%%%%%%%%%%%%%%%%%%%%%%%%%%%%%%%%%%%%%%%%%%%%%%%%%%%%
\acknowledgments     %>>>> equivalent to \section*{ACKNOWLEDGMENTS}       
 
J. Mazoyer is grateful to the CNES and Astrium (Toulouse, France) for supporting his PhD fellowship. The DM study at LAM was funded by CNES (Toulouse, France).

%%%%%%%%%%%%%%%%%%%%%%%%%%%%%%%%%%%%%%%%%%%%%%%%%%%%%%%%%%%%%
%%%%% References %%%%%

\bibliography{biblio_SPIE_montreal}   %>>>> bibliography data in report.bib

\begin{thebibliography}{1}

\bibitem{Bifano11}
Bifano, T., ``Adaptive imaging: {MEMS} deformable mirrors,'' {\em Nat
  Photon}~{\bf 5},  21--23 (Jan. 2011).

\bibitem{Liotard05}
{Liotard}, A., {Muratet}, S., {Zamkotsian}, F., and {Fourniols}, J.-Y.,
  ``Static and dynamic microdeformable mirror characterization by
  phase-shifting and time-averaged interferometry,'' in [{\em Society of
  Photo-Optical Instrumentation Engineers (SPIE) Conference
  Series}{\nolinebreak\hspace{0.1em}]},  {\em Society of Photo-Optical
  Instrumentation Engineers (SPIE) ConferenceSeries} {\bf 5716},  207--217
  (2005).

\bibitem{Hariharan87}
Hariharan, P., Oreb, B.~F., and Eiju, T., ``Digital phase-shifting
  interferometry: a simple error-compensating phase calculation algorithm,''
  {\em Appl. Opt.}~{\bf 26},  2504--2506 (Jul 1987).

\bibitem{Malacara07}
Malacara, D.,  [{\em Optical Shop Testing}{\nolinebreak\hspace{0.1em}]}, John
  Wiley \& Sons (Aug. 2007).

\bibitem{Huang08}
{Huang}, L., {Rao}, C., and {Jiang}, W., ``Modified gaussian influence function
  of deformable mirror actuators,'' {\em Optics Express}~{\bf 16},  108--114
  (Jan 2008).

\bibitem{Mazoyer13}
{Mazoyer}, J., {Baudoz}, P., {Galicher}, R., {Mas}, M., and {Rousset}, G.,
  ``Estimation and correction of wavefront aberrations using the self-coherent
  camera: laboratory results,'' {\em Astronomy and astrophysics}~{\bf 557},  A9
  (Sept. 2013).

\bibitem{Galicher14}
{Galicher}, R., {Delorme}, J.-R., {Baudoz}, P., {Mazoyer}, J., and {Rousset},
  G., ``High contrast imaging on the thd bench: progress and upgrades,'' in
  [{\em Society of Photo-Optical Instrumentation Engineers (SPIE) Conference
  Series}{\nolinebreak\hspace{0.1em}]},  {\em Society of Photo-Optical
  Instrumentation Engineers (SPIE) ConferenceSeries} (2014).

\bibitem{Delorme14}
{Delorme}, J.-R., {Galicher}, R., {Baudoz}, P., {Mazoyer}, J., and {Rousset},
  G., ``High contrast imaging in white light with a self-coherent camera,'' in
  [{\em Society of Photo-Optical Instrumentation Engineers (SPIE) Conference
  Series}{\nolinebreak\hspace{0.1em}]},  {\em Society of Photo-Optical
  Instrumentation Engineers (SPIE) ConferenceSeries} (2014).

\end{thebibliography}
\bibliographystyle{spiebib}   %>>>> makes bibtex use spiebib.bst

\end{document}